\documentclass[letter,11pt]{article}
\usepackage{jheppub}
\usepackage{multirow}

\def\epem{e^+e^-}
\def\abinv{\mathrm{ab}^{-1}}
\def\fbinv{\mathrm{fb}^{-1}}
\def\errors{\delta_{ij}}

\title{Prospects for the determination of the top-quark Yukawa
  coupling at future $\epem$ colliders}

\author[a]{Stefano Boselli,}
\author[b]{Ross Hunter,}
\author[a]{Alexander Mitov}

\affiliation[a]{Cavendish Laboratory, University of Cambridge, Cambridge CB3 0HE, UK}
\affiliation[b]{Department of Physics, University of Warwick, Coventry CV4 7AL, UK}

\note{Preprint: Cavendish-HEP-18/07}

\abstract{We estimate the sensitivity to the top-quark Yukawa coupling $y_t$ at future $\epem$ colliders. We go beyond the standard approach that focuses on $t\bar t h$ production and consider final states with a Higgs boson but not top quarks. The sensitivity to $y_t$ in such processes comes from the coupling of the Higgs boson to top quarks in loops. Such final states can be produced in significant numbers at center-of-mass energies that will be accessible by all proposed $\epem$ colliders. In a simplified theoretical framework to parametrise deviations from the Standard Model, we find that at FCC-$ee$ and CEPC operating at $\sqrt{s}=240$ GeV, $y_t$ could potentially be measured with precision better than 1\%. For CLIC and ILC the extraction of $y_t$ could be improved by a factor of about 2 and 7 respectively, compared to its extraction from just $t\bar t h$ final states.}

\begin{document} 
\maketitle
\flushbottom

\section{Introduction}\label{sec:intro}

The high-precision determination of the Higgs boson couplings is one
of the major tasks facing High Energy
Physics~\cite{Klute:2013cx,Mariotti:2016owy}. Despite the wealth of
information already delivered by the LHC~\cite{deFlorian:2016spz}, and
expected from its future high-luminosity operation, a truly
detailed study of the Higgs sector will likely be possible only at a
future $\epem$ collider.

The top-quark Yukawa
coupling $y_t$ dominates the renormalization group evolution of the
Higgs potential at high energy scales (see, for instance,
refs.~\cite{Degrassi:2012ry, Alekhin:2012py,Buttazzo:2013uya}). 
Therefore $y_t$ is among the main drivers of SM predictions at very high energies and often
dominates in, amongst others, studies of the self-consistency of the SM
at GUT-scale energies and in searches for physics beyond the SM.

The ideal way for measuring $y_t$ would be through $t\bar t h$ final
states since they allow cleaner interpretation of the measurement in
terms of $y_t$. However, to produce such a final state a center-of-mass (c.m.) energy of at least $500$ GeV is required.
From all proposed $\epem$ colliders - which we detail in 
sec.~\ref{sec:approach} below - only the Compact Linear Collider (CLIC) 
and the International Linear Collider (ILC) will be capable of 
achieving such c.m. energies. Preliminary studies concerning $t\bar t h$ final states suggest~\cite{Abramowicz:2016zbo, Barklow:2015tja, Fujii:2015jha} that CLIC and ILC will be 
able to measure $y_t$ with a precision of about 4--5\%. Given the importance of the top-quark Yukawa
coupling such ultimate precision is not entirely satisfactory. Indeed,
it can be contrasted to the expected 10\% precision~\cite{CMS:2013xfa}
from the HL-LHC and the 1\% precision expected at a future
100 TeV hadron collider \cite{Plehn:2015cta}.

Given the central importance of the top-quark Yukawa coupling as well
as the seemingly puzzling fact that future $\epem$ colliders may not
be able to measure it better than a future hadron collider, in this
work we set ourselves the goal of addressing the following question:
{\it what is the ultimate precision with which $y_t$ can be measured at future $\epem$ colliders}? 
  
To answer this question we explore a new approach for the determination of the top-quark Yukawa coupling, utilizing loop-induced Higgs production and decay processes, in a simple version of the so-called $\kappa$-framework~\cite{LHCHiggsCrossSectionWorkingGroup:2012nn}. Loop-induced processes have an advantage in that they potentially allow for a precise determination of $y_t$ at colliders such as FCC-$ee$ or CEPC which are designed to operate at c.m. energies below the $t\bar t h$ threshold. Furthermore, even at colliders that can produce $t\bar t h$ final states, measurements at different 
c.m. energies could be combined in order to derive more precise determination of $y_t$ than from $t\bar t h$ final states alone. Such indirect approaches are already being pursued in, for example, the determination of the Higgs self-interaction at the LHC~\cite{Degrassi:2016wml, Maltoni:2017ims, Maltoni:2018ttu,Bizon:2016wgr, DiVita:2017vrr, DiVita:2017eyz}. 

This work is organized as follows: in sec.~\ref{sec:approach} we provide a general overview of future $\epem$ colliders and the Higgs processes considered in this paper. In sec.~\ref{sec:fit} we describe our approach for determining $y_t$ and detail the fitting procedure adopted in our analysis. Our numerical results are presented in sec.~\ref{sec:results}. In sec.~\ref{sec:limitations} we discuss the limitations of our study and possible future extensions. Our conclusions are summarized in sec.~\ref{sec:conclusions}.

\section{Higgs production and decay processes at $\epem$ colliders}\label{sec:approach}

A number of future lepton colliders have been proposed over the years:
\footnote{Since this study was conducted and this publication prepared, FCC-$ee$, CEPC and CLIC have all produced updates to their operational baselines. The potential consequences for our analysis are discussed in sec. \ref{sec:noteadded}.}
\begin{itemize}
\item The Circular Electron Positron Collider (CEPC) in China would
collect 5 $\abinv$ of integrated luminosity at 240
GeV~\cite{CEPC-SPPCStudyGroup:2015csa, CEPC-SPPCStudyGroup:2015esa, Mo:2015mza}. 
A run at 350 GeV could also be envisioned;
\item The Future Circular Collider with $\epem$ (FCC-$ee$) is a 
high-luminosity, high-precision circular collider envisioned in a 
new 80-100 km tunnel at CERN~\cite{Gomez-Ceballos:2013zzn,dEnterria:2016sca}. 
The FCC-$ee$  aims at collecting multi-$\abinv$ integrated luminosity at $\sqrt{s} =$ 90, 160, 240, and 350 GeV. 
In particular, 10 $\abinv$ of data would be  collected at 240 GeV and 2.6 $\abinv$ at 350 GeV;
\item The Compact Linear Collider (CLIC) at CERN would collect 100 
$\fbinv$at the top threshold, 500 $\fbinv$ at 380 GeV, 1.5 $\abinv$ 
at 1.5 TeV, and 3 $\abinv$ at 3 TeV~\cite{CLIC:2016zwp}. 
The study of Higgs measurements at CLIC~\cite{Abramowicz:2016zbo} assumes a different scenario: 
500 $\fbinv$ at 350 GeV, 1.5 $\abinv$ at 1.4 TeV and 2 $\abinv$ at 3 TeV. 
In the following we assume this scenario, for which an accurate estimate of the relevant experimental uncertainties is available;
\item The International Linear Collider (ILC) is a proposed linear
$\epem$ collider to run in the energy range
between 200 GeV and 500 GeV~\cite{Barklow:2015tja, Fujii:2015jha}.
Here we follow a scenario with three energy stages: 
250, 350, and 500 GeV, and accumulated luminosity of, respectively, 2
$\abinv$, 200 $\fbinv$ and 4 $\abinv$. The 350 GeV stage of the ILC 
is mainly intended for the determination of the top quark mass 
from the $t\bar{t}$ threshold. This stage will not be considered in our study due to its low luminosity.
\end{itemize} 
\begin{figure}[t]
\includegraphics[width = 1.06\textwidth]{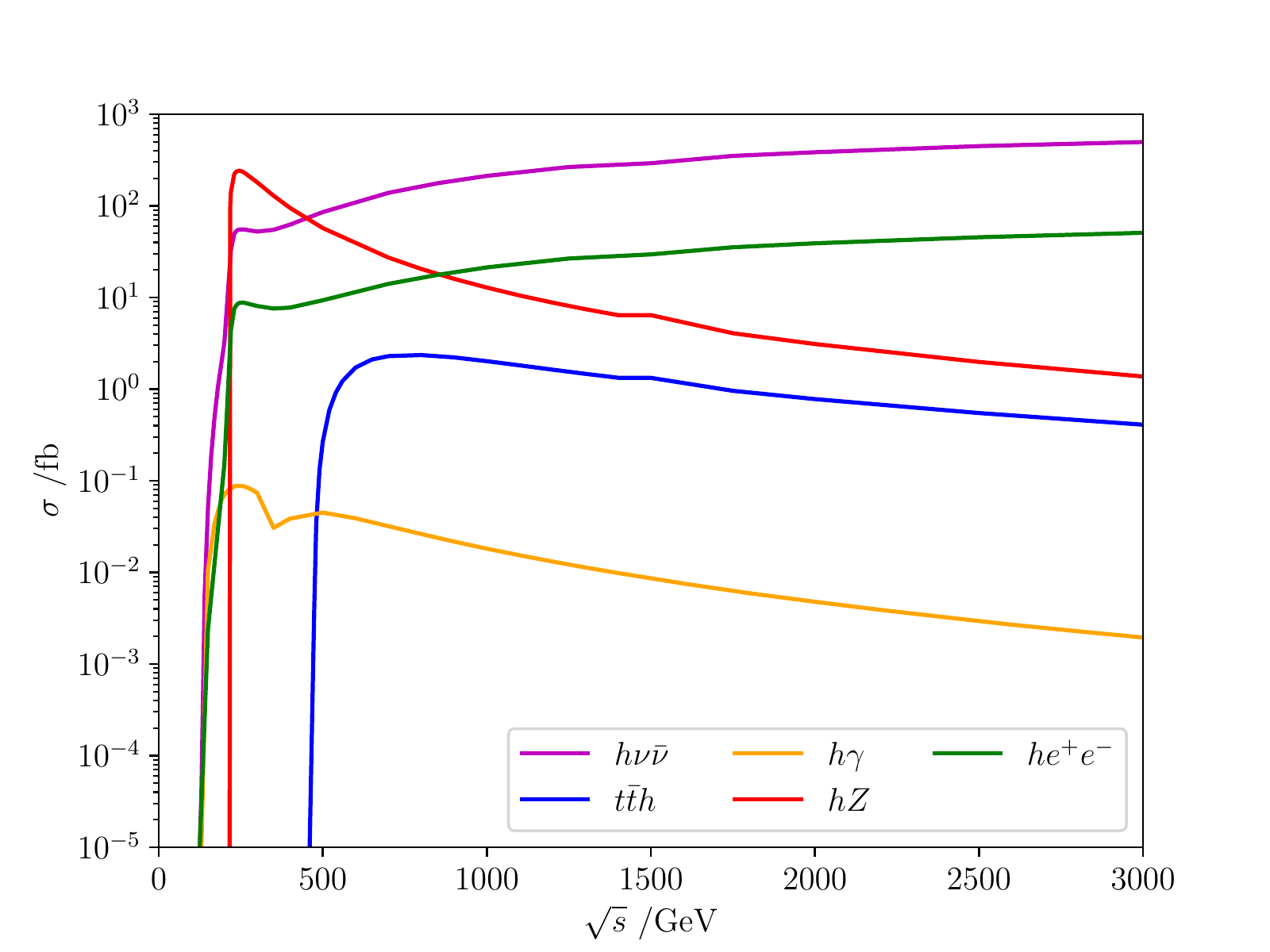}
\caption{LO cross-section for the single Higgs production channels at future lepton colliders described in Sec.~\ref{sec:approach}. The process labeled $hZ$ includes all decays of the $Z$ boson.}
\label{fig:scan}
\end{figure}
The dominant single-Higgs production mechanisms at the above future colliders are the $s$-channel Higgstrahlung process $\epem \to hZ$ and the $t$-channel charged vector-boson fusion (VBF) process resulting in $h \nu \bar{\nu}$ final states. The relative importance of these two processes depends on the c.m. energy; the Higgstrahlung process dominates around 240-250 GeV while the VBF cross section takes over around 500 GeV. At even higher energies, the neutral VBF process $\epem\to he^-e^+$ also becomes significant. One should keep in mind, however, that the separation of the various processes is not unambiguous once the $Z$ decays have been taken into account. In particular, Higgstrahlung with $Z$ decaying to neutrinos (electrons) yields the same final-state as the charged (neutral) vector boson fusion processes. This contamination is particularly relevant at 240 GeV where the inclusive $h \nu \bar{\nu}$ rate is dominated by Higgstrahlung. 

For this reason in fig.~\ref{fig:scan} we show the c.m. dependence of the computed at leading order (LO) inclusive cross-section for the final states described above. The $h \nu \bar{\nu}$ and $h \epem$ channels include the Higgstrahlung contribution; the process labeled $hZ$ on the other hand includes all $Z$ decay modes. The same applies to the results in tables~\ref{tab:xs1},\ref{tab:xs2} below. We note, however, that in table~\ref{tab:1s} the process labeled $hZ$ is not inclusive in the $Z$ decay and includes only certain $Z$ decay modes which are specific to the analyses referenced in that table.

In this study we are interested in processes which 
are sensitive to a non-vanishing anomalous contribution to $y_t$. 
For this reason, in fig.~\ref{fig:scan} we show two more Higgs production processes: 
$t\bar{t}h$, which is usually considered as the only available channel for the 
extraction of  $y_t$, and the loop-induced process $h\gamma$ 
which is one of the inputs to our analysis.

To get an overall impression about the potential of the various Higgs production modes, in tables~\ref{tab:xs1} and~\ref{tab:xs2} we show the expected number of events for each run of the future colliders described above. The expected numbers of events are derived by multiplying the inclusive cross-sections with the corresponding luminosities shown in tables~\ref{tab:xs1} and~\ref{tab:xs2}.

As far as Higgs decays are concerned, the loop-induced processes $h\to gg$ and $h\to\gamma\gamma$ are both sensitive to $y_t$ and will be considered in the following. The Higgs decay to gluons is generated by massive quarks in the loops with the top-quark being the dominant contribution. In the $m_t \to \infty$ limit this coupling is known with next-to-next-to-next to leading order  (${\rm N}^3{\rm LO}$) QCD accuracy~\cite{Baikov:2006ch}. In contrast, the Higgs decay to photons (as well as $h\gamma$ production) has a dominant contributions from loops involving gauge bosons which results in a reduced sensitivity to $y_t$ compared to $h\to gg$. 
\begin{table}
	\begin{center}
		\begin{tabular}{l | c | c | c  }
                		&  \multicolumn{2}{ c |}{FCC-ee} & CEPC \\
			\hline
			$\sqrt{s}$ (GeV) & 240 & 350 & 240  \\
			$\mathcal{L}_\mathrm{int.}$ ($\mathrm{fb}^{-1}$) & 
			$1.0 \cdot 10^4$& $2.6 \cdot 10^3$ & $5.0 \cdot 10^3$ \\
			\hline
			$\sigma_{hZ}$ (fb) & 240 & 130 & 240 \\
			$\mathcal{N}_{hZ}$ &$2.4 \cdot 10^6$ & $3.38 \cdot 10^5$ &
			$1.2 \cdot 10^6$ \\ 
			\hline
			$\sigma_{\nu\bar{\nu}h}$ (fb) & 54.4 & 54.7 
			& 54.4 \\
			$\mathcal{N}_{\nu\bar{\nu}h}$&$5.44 \cdot 10^5$ 
			&$1.42 \cdot 10^5$&$2.72 \cdot 10^5$ \\ 
			\hline
			$\sigma_{eeh}$ (fb) & 7.9 & 7.13 & 7.9 \\
			$\mathcal{N}_{eeh}$ & $7.9 \cdot 10^4 $ & $1.85 \cdot 10^4$ &
			$3.95 \cdot 10^4$ \\ 
			\hline
			$\sigma_{h\gamma}$ (fb) & $8.96\cdot 10^{-2}$ 
			& $3.18 \cdot 10^{-2}$ & $8.96\cdot 10^{-2}$ \\
			$\mathcal{N}_{h\gamma}$ & 896 & 82 & 448 \\ 
			\hline			
		\end{tabular}
	\end{center} 
	\caption{Inclusive LO cross-sections and numbers of expected events 
	for the main Higgs production modes at FCC-$ee$ and CEPC. 
	The process labeled $hZ$ includes all decays of the $Z$ boson.}
	\label{tab:xs1}
\end{table}
\begin{table}
	\begin{center}
		\begin{tabular}{l | c | c | c | c | c }
			& \multicolumn{3}{ c |}{CLIC} &  \multicolumn{2}{ c }{ILC} \\
			\hline
			$\sqrt{s}$ (GeV)& 350 & 1400 & 3000 & 250 & 500 \\
			$\mathcal{L}_\mathrm{int.}$ ($\mathrm{fb}^{-1}$) & 
			$5.0\cdot 10^2$ & $1.5 \cdot 10^3$ & $2.0 \cdot 10^3$ &
			$2.0 \cdot 10^3$ & $4.0 \cdot 10^3$\\
			\hline
			$\sigma_{hZ}$ (fb) & 130 & 6.42 & 1.37  & 240 & 57.2 \\
			$\mathcal{N}_{hZ}$ & $6.50 \cdot 10^4$ & $9.6 \cdot 10^3$ 
			& $2.74 \cdot 10^3$  & $4.80 \cdot 10^5$ & $2.29 \cdot 10^5$ \\ 
			\hline
			$\sigma_{\nu\bar{\nu}h}$ (fb) & 54.4 & 293 & 498 & 55.0 & 85.2    \\
			$\mathcal{N}_{\nu\bar{\nu}h}$ &$2.73 \cdot 10^4$&
			$4.39 \cdot 10^5$&$9.96 \cdot 10^5$ & $1.10 \cdot 10^5$ & 
			$3.41 \cdot 10^5$   \\ 
			\hline
			$\sigma_{eeh}$ (fb) & 7.13 &  28.3 & 49.1 & 8.2 & 8.7   \\
			$\mathcal{N}_{eeh}$ & $3.56 \cdot 10^3$ & 
			$4.24\cdot 10^4$& $9.82 \cdot 10^4$ & $1.64 \cdot 10^4$ & 
			$3.48 \cdot 10^4$   \\ 
			\hline
			$\sigma_{t\bar{t}h}$ (fb) & - & 1.33 & 0.41 & - & 0.27  \\
			$\mathcal{N}_{t\bar{t}h}$ & - & 1995 & 820 & - & $1.08 \cdot 10^3$ \\ 
			\hline
			$\sigma_{h\gamma}$ (fb) & $3.18 \cdot 10^{-2}$ 
			& $1.20 \cdot 10^{-2}$ & $3.08 \cdot 10^{-3}$ & 
			$8.97 \cdot 10^{-2}$ & $4.74 \cdot 10^{-2}$ \\
			$\mathcal{N}_{h\gamma}$ & 16 & 18 & 6 & 179 & 189 \\ 
			\hline			
		\end{tabular}
	\end{center} 
	\caption{As in table~\ref{tab:xs1} but for CLIC and ILC.}
	\label{tab:xs2}
\end{table}

\newpage

\section{Our approach for determining $y_t$}\label{sec:fit}

The problem of determining $y_t$ at a given run of a hypothetical future collider is formulated in terms of the new physics contribution $\Delta y_t$ to the top-quark Yukawa coupling $y_t$

\begin{equation}
y_t = y_t^\mathrm{SM} + \Delta y_t\,,
\label{eq:resc}
\end{equation}
where we assume that this is the only source of deviation from the SM. This assumption is discussed in sec. \ref{sec:limitations}.  

The main limitation of an analysis restricted to $t\bar th$ data is that it requires a c.m. energy of at least 500 GeV. To circumvent this limitation, we also consider $\epem$ observables which are {\it indirectly} sensitive to $y_t$ and, at the same time, have sufficiently large number of expected events. The $y_t$ dependence in such processes originates in the Higgs boson coupling to top quarks in loops, either in the production or in the decay of the Higgs boson
\footnote{We will not consider final states with top quarks but no Higgs. These final states are included in the $t\bar t$ threshold scan studies~\cite{Beneke:2015lwa, Beneke:2016kkb, Beneke:2017rdn}.}. 
Thus, in addition to $t\bar t h$ production, we consider the Higgs decays to gluons and photons as well as Higgs production in association with a hard photon. The idea is to exploit the $y_t$ dependence of single Higgs processes with the added benefit that these processes are accessible at all c.m energies.

In order to constrain $\Delta y_t$ we define a global $\chi^2$ for each run of the future colliders described in the previous section
\begin{equation}
\chi^2 (\Delta y_t) =  \sum_{i=1}^{N_p}\sum_{j=1}^{N_d} \frac{\left[\mu_{ij}\left(\Delta y_t\right) - 1\right]^2}{\errors^2} \,,
\label{eqn:chi}
\end{equation}
with $N_p$ and $N_d$ being, respectively, the number of available production and decay channels. The sums in eq.~(\ref{eqn:chi}) include only the processes for which $\errors$ values are explicitly shown in table~\ref{tab:1s}. 

The degrees of freedom of the $\chi^2$  in eq.~(\ref{eqn:chi}) are the signal-strengths $\mu_{ij}$ of all Higgs boson processes which are sensitive to a non-vanishing value of $\Delta y_t$ and which can be measured with a sufficient precision. The signal-strength $\mu_{ij}$ for a generic Higgs production mode $i$ and decay channel $j$ can be written in the narrow-width approximation as
\begin{equation}
\mu_{ij} = 
\left(\frac{\sigma_i}{\sigma_i^\mathrm{SM}}\right) 
\left(\frac{\Gamma_j}{\Gamma_j^\mathrm{SM}}\right) 
\left(\frac{\Gamma_h}{\Gamma_h^{\mathrm{SM}}}\right)^{-1}\,.
\label{eqn:muxy}
\end{equation}
In eq.~(\ref{eqn:muxy}) $\Gamma_h$ is the total Higgs width and $\sigma_i$ and $\Gamma_j$ are the corresponding production cross-section and partial decay width. Due to the small number of expected $t\bar th$ and $h\gamma$ events, these two production channels are included with only the dominant $h\to b\bar b$ decay mode. 

The one-sigma uncertainties $\errors$ appearing in eq.~(\ref{eqn:chi}) and listed in table~\ref{tab:1s} are taken from the literature~\cite{Abramowicz:2016zbo,Gomez-Ceballos:2013zzn, CEPC-SPPCStudyGroup:2015csa, Durieux:2017rsg}. Their values have been derived in a realistic framework that accounts for acceptance cuts, background contributions and detector simulation for the reconstruction of the final state. To the best of our knowledge such studies are not available for the $h\gamma$ process. For this reason the one-sigma uncertainties for this channel have been derived by calculating the Poissonian error and have to be considered as optimistic estimates of the total uncertainties. 
\begin{table}
	\begin{center}
		\begin{tabular}{c | c | c | c | c | c | c | c | c }
			\multirow{2}{*}{Collider} & 
			\multirow{2}{*}{$\sqrt{s}$ (GeV)} & 
			\multirow{2}{*}{$\mathcal{L}$
                          $\left(\mathrm{fb}^{-1}\right)$} &
			\multicolumn{2}{ c |}{$h \to g g$} &  
			\multicolumn{2}{ c |}{$h \to \gamma \gamma$} &
			\multicolumn{2}{ c }{$h \to b\bar{b}$}  \\
                        \cline{4-9}      		
			& & & $hZ$ & $\nu\bar{\nu}h$ 
			& $hZ$ & $\nu\bar{\nu}h$ & $h\gamma$  & $t\bar{t}h$ \\
			\hline			           
                	\multirow{2}{*}{FCC-{\it ee}} & 240 &$1.0\cdot 10^4$& 1.4\% & - 
			& 3.0\% & - & 4.4\% & - \\
               		& 350 & $2.6\cdot 10^3$& 3.1\% & 4.7\% 
			& 14\%& 21\% & 14\% & - \\
			\hline
                	CEPC & 240 & $5.0\cdot 10^3$ & 1.2\% & - 
			& 9.0\% & - & 6.2\% & - \\
			\hline
		        \multirow{3}{*}{CLIC}& 350 & $5.0\cdot 10^2$ &
                        6.1\% & 10\%
		        & - & - & - & -  \\                 			
		        & 1400 & $1.5\cdot 10^3$ & - & 5.0\% 
		        & - & 15\% & - & 8.0\% \\  
		        & 3000 & $2.0\cdot 10^3$ &
                         - & 4.3\% & - & 10\% & - & 12.5\%  \\    
			\hline
			\multirow{2}{*}{ILC}& 250 & $2.0\cdot 10^3$ &
                        2.5\% & - & 12\%
			& - & 10\% & -    \\                 			
		        & 500 & $4.0 \cdot 10^3$ & 3.9\% & 1.4\% &
                        12\% & 6.7\% & 9.8\%  & 9.9\%
		\end{tabular}			
		\caption{The estimated one-sigma uncertainties 
		$\errors$ used in eq.~(\ref{eqn:chi}).
                 The ones for CLIC, CEPC and FCC-{\it ee} at $\sqrt{s}=240$ GeV are taken from
                  refs.~\cite{Abramowicz:2016zbo,Gomez-Ceballos:2013zzn, 
                  CEPC-SPPCStudyGroup:2015csa}, respectively. 
                  The ones for ILC and FCC-{\it ee} at $\sqrt{s}=350$ GeV are from ref.~\cite{Durieux:2017rsg}. 
                  The $\sqrt{s}=3$ TeV CLIC $t\bar{t}h$ result is derived
                  by extrapolating the 1.4 TeV one with the corresponding number of events.
                  The statistical uncertainties for $h\gamma$ production are derived from the expected number of events
                  reported in tables~\ref{tab:xs1} and~\ref{tab:xs2}. 
                  The process labeled $hZ$ includes selected $Z$ decays, and their content is specific to each analysis referenced in this table.}
		\label{tab:1s}
	\end{center}
\end{table}

The analytic expressions for the $h\gamma$ and $t\bar{t}h$ signal-strengths, as functions of $\Delta y_t$, read
\begin{eqnarray}
	\mu_{h\gamma}
	  \begin{pmatrix}
          \sqrt{s} = 240 \, \mathrm{GeV} \\
          \sqrt{s} = 250 \, \mathrm{GeV} \\
          \sqrt{s} = 350\, \mathrm{GeV} \\
          \sqrt{s} = 500\, \mathrm{GeV}
        \end{pmatrix} &=& 
	\frac{\sigma_{h\gamma}}{\sigma^{\mathrm{SM}}_{h\gamma}} = 
	1 - \begin{pmatrix}
          0.43 \\
          0.45 \\
          0.73 \\
          0.13
        \end{pmatrix} \Delta y_t \label{eq:hahg} \\
	\mu_{t\bar{t}h}
        \begin{pmatrix}
          \sqrt{s} = 500 \, \mathrm{GeV} \\
          \sqrt{s} = 1400\, \mathrm{GeV} \\
          \sqrt{s} = 3000\, \mathrm{GeV}
        \end{pmatrix} &=& 
	\frac{\sigma_{t\bar{t}h}}{\sigma^{\mathrm{SM}}_{t\bar{t}h}} = 
	1 + \begin{pmatrix}
          1.99 \\
          1.83 \\
          1.71
        \end{pmatrix} \Delta y_t,
	\label{eq:ha}
\end{eqnarray}

In the calculation of the above expressions we do not include corrections beyond LO. Such higher-order effects have been studied in ref.~\cite{Abramowicz:2016zbo} for the 1.4 TeV run of CLIC. These corrections result in a relatively small shift in the corresponding coefficient in eq.~(\ref{eq:ha}) from 1.83 to 1.89. In turn, this slightly increases the $y_t$ precision in the $t\bar th$ channel.

For the two loop-induced Higgs decay processes we get
\begin{eqnarray}
	\mu_{h\to g g} &=&  
	\frac{\Gamma_{h\to gg}} {\Gamma_{h\to gg}^\mathrm{SM}} =
	1 + 2  \Delta y_t\,,  \label{eq:widthhgg}\\
	\mu_{h\to \gamma \gamma} &=& 
	\frac{\Gamma_{h\to \gamma\gamma}}
	{\Gamma_{h\to \gamma\gamma}^\mathrm{SM}} =
	1 -0.56  \Delta y_t\,.
	\label{eq:width}
\end{eqnarray}

All computations in this work have been carried out in the $G_\mu$ input scheme with the help of the \texttt{Madgraph5\_aMC@NLO\_v2.6.1} code~\cite{Alwall:2014hca}. Eqs.~(\ref{eq:hahg}--\ref{eq:width}) have been derived in the following way: we first compute the corresponding cross-sections and decay widths for a number of different values of $\Delta y_t$ and then fit the resulting expressions for $\mu_{ij}$  with a parabola. Finally, we take its linear approximation for small values of $\Delta y_t$. In deriving $\mu_{h\to g g}$ the bottom quark contribution in the loop has been neglected.

\section{Results}\label{sec:results}

\begin{figure}[t]
\includegraphics[width = 1.05\textwidth,trim=0 5mm 0 5mm]{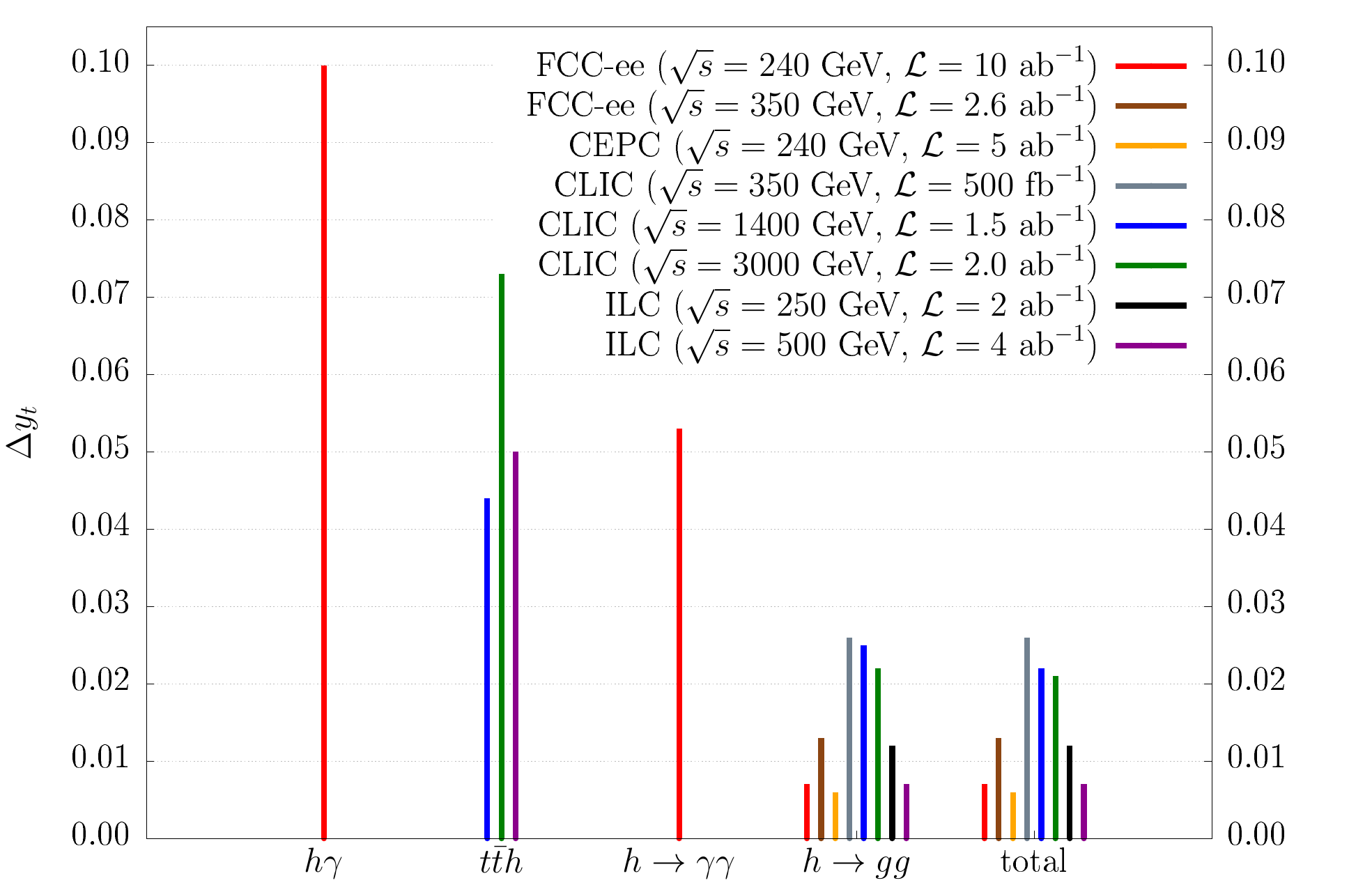}
\label{fig:mu}
\caption{68\% CL boundaries on $\Delta y_t$ for different runs and
  processes. In the last column we report the results of the global
  $\chi^2$ analysis described in sec.~\ref{sec:fit}.}
\label{fig:fit}
\end{figure}

Our main results, namely, the 68\% CL constraints following from eq.~(\ref{eqn:chi}), are displayed
in table~\ref{tab:yt} and in fig.~\ref{fig:fit}. We report results for
the following scenarios: the FCC-$ee$ runs at c.m. energies of
$240$ GeV and $350$ GeV, the CEPC run at $240$ GeV, the three CLIC runs at $350$ GeV, $1.4$ TeV and $3$ TeV 
and the $250$ GeV and $500$ GeV runs of the ILC. 
\begin{table}
	\begin{center}
		\begin{tabular}{ c | c | c | c | c | c | c | c}
			Collider & 
			$\sqrt{s}$ (GeV) & 
			$\mathcal{L}$ $\left(\mathrm{fb}^{-1}\right)$ & 
			$h \to g g$ &  
			$h \to \gamma \gamma$ &
			$h\gamma$  & 
			$t\bar{t}h$ & 
			total \\
			\hline
			\multirow{2}{*}{FCC-{\it ee}} & 240 &$1.0\cdot 10^4$ & 0.7\% 
			&  5.3\% & 10\% & - & 0.7\%  \\
			& 350 &$2.6\cdot 10^3$& 1.3\% 
			& 21\% & 19\% & - & 1.3\% \\
                		\hline
                		CEPC & 240 & $5.0\cdot 10^3$ & 0.6\%
                                & 16\% & 14\% & - & 0.6\%  \\
			\hline
		        \multirow{3}{*}{CLIC}& 350 & $5.0\cdot 10^2$ & 2.6\%& 
		        - & - & - & 2.6\% \\               	
		        & 1400 & $1.5\cdot 10^3$ & 2.5\% & 27\% & - &
                        4.4\% & 2.2\% \\
                        & 3000 & $2.0\cdot 10^3$ & 2.2\% & 18\% & - & 7.3\% & 2.1\% \\
                        \hline
                        \multirow{2}{*}{ILC}& 250 & $2.0\cdot 10^3$
                        &1.2\% & 21\% & 23\% & - & 1.2\% \\               	
		        & 500 & $4.0\cdot 10^3$ & 0.7\%  & 10\% & 75\%
                        & 5.0\% & 0.7\%
                        \end{tabular}
		\caption{68\% CL boundaries on $\Delta y_t$ for
                  different runs and processes. In the last column we
                  report the results of the global $\chi^2$ analysis
                  described in sec.~\ref{sec:results}.}
		\label{tab:yt}
	\end{center}
\end{table}

From table~\ref{tab:yt} and fig.~\ref{fig:fit} we conclude that the decay process $h \to gg$ is a strong potential candidate for precise determination of $y_t$. Combining the high $y_t$ sensitivity of $h \to gg$ seen in eq.~(\ref{eq:widthhgg}) with the high luminosities at the 240 GeV FCC-$ee$ and CEPC runs and at the 500 GeV ILC run, one may potentially be able to determine $\Delta y_t$ with uncertainty of 0.6--0.7\%. 

The potential of $h \to \gamma\gamma$ for a precise determination of $y_t$ is much smaller than $h\to gg$. Despite the low cross-section of $\epem\to h\gamma$ (see fig.~\ref{fig:scan}), this loop-induced process allows access to $y_t$ at both FCC-$ee$ and CEPC. While not directly competitive with $h\to gg$, this additional $y_t$ sensitivity is on par with the one expected at HL-LHC and may be useful for disentangling Wilson coefficients in a more refined Effective Field Theory (EFT) approach (see sec.~\ref{sec:limitations} for further details and ref.~\cite{Brivio:2017vri} for a recent review). 

As far as CLIC is concerned, its 350 GeV run allows $y_t$ to be determined from purely loop-induced processes with precision of about 2.6\%. 
At higher CLIC energies the precision in the $y_t$ determination from loop-induced processes 
is significantly larger than the one expected from the standard $t\bar t h$-based approach. Our estimates show that by combining the
extraction of $y_t$ from $t\bar th$ with that from loop-induced final states one can
reach $y_t$-precision of about 2.1--2.2\% at both the $\sqrt{s}=1.4$ TeV and $\sqrt{s}=3.0$ TeV CLIC runs. This is 2-3 times better than the precision expected from purely $t\bar th$ final states. 
 
From table~\ref{tab:yt} and fig.~\ref{fig:fit} we also conclude that for all collider runs considered by us, loop-induced processes (mostly $h\to gg$) could potentially lead to significantly more precise determination of $y_t$ compared to $t\bar th$ final states. 

\section{Limitations of the present study and possible further improvements}\label{sec:limitations}

The precision in the various $y_t$ determinations estimated in the previous
section are based on a number of assumptions and approximations. We
discuss them in turn.

For most processes and colliders we have used existing studies for
Higgs production and decay. An exception is the process $\epem\to h\gamma$~\cite{Barroso:1985et, Abbasabadi:1995rc, Djouadi:1996ws, Sang:2017vph}.
In this work we have computed this process at LO in the SM with the help of
\texttt{Madgraph5\_aMC@NLO\_v2.6.1}, i.e. accounting fully for the loop in the Higgs production process
\footnote{The rational terms in the relevant \texttt{Madgraph5\_aMC@NLO\_v2.6.1} model file are written in terms of the top-quark mass. They need to be recast in terms of $y_t$ in order to obtain the correct $y_t$ dependence of the $h\gamma$ cross-section.}. 
No detector simulation, efficiency estimates or realistic estimate of
backgrounds and systematic effects have been performed by us for this
process. For this reason the purely statistical errors derived by us are, likely, optimistic.

In all our estimates we assume that the top-quark mass $m_t$ is
perfectly known. While the future lepton colliders capable of reaching
the $t\bar t$ threshold of $\sqrt{s}= 350$ GeV should be able to
measure $m_t$ with excellent precision of about 50
MeV~\cite{Beneke:2015lwa, Beneke:2016kkb, Beneke:2017rdn,
  Penin:2014zaa, Beneke:2015kwa, Beneke:2016cbu}, the timing of such
measurements may be an issue. For example, a $t\bar t$ threshold scan
may come only after the measurements at 240 GeV have been
performed. In any case one can either reanalyze the lower energy
measurement in light of new $m_t$ measurements or utilize then-available HL-LHC
measurements. Indeed, the ${\cal O}(300~{\rm MeV})$ uncertainty on $m_t$ expected from HL-LHC will already be sufficiently precise to be a subdominant effect for even the most precise expected determination of $y_t$.

An important caveat for our study is the interpretation of the
possible measurements in terms of uncertainty on $y_t$
alone. As we already mentioned in sec.~\ref{sec:fit}, we work 
in a simple extension of the SM where $\Delta y_t$ is the only 
possible anomalous coupling. In effect this allows us to simply trade any uncertainty in the
measurement for an uncertainty in $y_t$. In reality such an assumption
is not well motivated since the assumption of non-vanishing $\Delta y_t$
implies physics beyond the SM, and once such an assumption is made
then there is no good justification for assuming a single source of deviation from SM. In this
sense, ideally, one would like to treat the problem of the $y_t$
determination within the $\kappa$-framework or a
full-blown EFT approach (see
refs.~\cite{Elias-Miro:2013mua, Jenkins:2013zja, Pomarol:2013zra, Jenkins:2013wua, Alonso:2013hga, Ellis:2014dva,
Falkowski:2015fla, Butter:2016cvz, Kobakhidze:2016mfx, Ellis:2014jta, Brivio:2017bnu, deBlas:2018tjm, Ellis:2018gqa, Vryonidou:2018eyv}
for global analyses in the LHC framework and
refs.~\cite{Durieux:2017rsg, Ellis:2015sca, Ellis:2017kfi} for EFT
analyses in the context of future lepton colliders).

While the introduction of many EFT couplings will reduce the
sensitivity to $y_t$ from the measurements we discuss, one should bear in mind that we have not assumed any prior knowledge on $y_t$ or any other EFT
coupling. In reality the LHC, especially after its
high luminosity phase, is expected to produce a significant set of
constraints on both $y_t$ and the EFT couplings that will
contribute to the extraction of $y_t$ at future lepton
colliders (a recent LHC update with HL-LHC projections for relevant operators can be found in ref.~\cite{Maltoni:2016yxb}). 
This will benefit the proposed extraction of $y_t$.   

Our indirect approach for the extraction of $y_t$ could also be applied/contrasted with the HL-LHC. 
A naive translation of the latest projections from a $\kappa$-framework study on Higgs physics at the HL-LHC~\cite{Cepeda:2019klc}
into our framework using $\kappa_g^2 = \Gamma_{h\to gg}/\Gamma_{h\to gg}^\mathrm{SM}$ 
and eq.~(\ref{eq:widthhgg}), yields $\Delta y_t \approx 2.5\%$. This is better than the expected direct $y_t$ determination from $t\bar th$ 
final states, but still around a factor of four worse than what some future $\epem$ colliders can achieve. 

Another tacit assumption made in our analysis is the perfect (or near
perfect) knowledge of the SM predictions for the processes under
consideration. Full NLO SM accuracy is now easily achievable for
non-loop-induced process thanks to automated tools like
\texttt{Madgraph5\_aMC@NLO\_v2.6.1},
\texttt{Sherpa}~\cite{Gleisberg:2008ta} and
\texttt{Whizard}~\cite{Moretti:2001zz,Kilian:2007gr}.  Full NLO SM
accuracy is also desired for the loop-induced processes discussed here. 
This requires the calculation of two-loop multiscale
amplitudes. A lot of progress in this direction has recently been achieved at
the LHC~\cite{Kudashkin:2017skd,Jones:2018hbb,Borowka:2016ypz,Borowka:2016ehy,Davies:2018ood,Borowka:2017idc,Bogner:2017xhp,Neumann:2018bsx,Borowka:2018dsa} 
making this a doable, albeit non-trivial, problem.

\section{Conclusions}\label{sec:conclusions}

We estimate the ultimate precision with which the top-Yukawa coupling $y_t$ can be extracted at the various proposed high-energy $\epem$ colliders, utilising a simplified version of the $\kappa$-framework. Our motivation for embarking on this study stems from the recognition that the traditional approach for extracting $y_t$ from $t\bar t h$ final states may be too restrictive: such an approach can only be realised at a couple of proposed colliders (CLIC and ILC) and in both cases results in somewhat limited precision of about 4--5\%. Such precision has to be viewed in the context of the high-luminosity LHC, which will precede any future $\epem$ machine, and where precision on $y_t$ of about 10\%, or even better, is expected.

To increase the scope for precise extraction of $y_t$ at future $\epem$ colliders, in this work we consider an alternative set of final states that are {\it indirectly} sensitive to $y_t$. The main advantage in considering loop-induced processes is that due to their large expected event yields and sensitivity to $y_t$, such processes can significantly increase the range of lepton colliders at which $y_t$ can be precisely determined. 

We find that potentially one could measure $y_t$ with precision of about 0.6\%  at the 240 GeV CEPC run. Similarly, the 240 GeV FCC-$ee$ and 500 GeV ILC runs have the potential for determining $y_t$ with precision of 0.7\%. Such high $y_t$ precision is driven mainly by the Higgs decay to gluons. Furthermore, the inclusion of $h\to gg$ data significantly increases the sensitivity to $y_t$ at higher c.m. energies. For example, at the 1.4 TeV CLIC run one can get an improvement by a factor of about two compared to $t\bar th$-only data.

The loop-induced Higgs decay $h\to\gamma\gamma$ is not as sensitive to $y_t$ but still offers a decent, better than 6\%, precision at the FCC-$ee$ collider. While not directly competitive with the $h\to gg$ decay, $h\to\gamma\gamma$ data could be useful for disentangling contributions from effective couplings and in some cases offers precision better than the one expected form the HL-LHC. 

Finally, we have identified the loop-induced associated process $\epem\to h\gamma$, with $h\to b\bar b$, which does not rely on loop-induced Higgs decays and could allow $y_t$ precision of about 10\%. Such a precision is comparable to the one expected from the HL-LHC.

\section{Note added}\label{sec:noteadded}

As mentioned in section \ref{sec:approach}, after our work was completed and made available on arXiv, updates to the baselines of FCC-$ee$, CEPC and CLIC appeared. The CEPC CDR~\cite{CEPCStudyGroup:2018ghi} reports a modest $\sim 10\%$ increase in integrated luminosity. In contrast, CLIC now plans to take more than double the integrated luminosity than quoted here, whereas the FCC-$ee$ data-taking expectations~\cite{Mangano:2018mur} have been roughly halved. Both have also made small changes to their planned centre-of-mass energies. Whilst these are large operational changes, none are predicted to make a significant effect on the $h\to gg$ process. The $h\to gg$ signal strength (eq.~(\ref{eq:widthhgg})) is not affected by $\sqrt{s}$, while the experimental error on extracting $\sigma \times \textrm{BR}(h\to gg)$ - as listed in table \ref{tab:1s} - is not predicted by FCC-$ee$ or CEPC to significantly change with their new baselines. Since we find that this process dominates the extraction of $y_t$ at all future lepton colliders, the broad conclusions we have drawn here remain unchanged. This is illustrated by an approximate scaling of our results performed in light of CLIC's new baseline, which modestly improves CLIC's reach to below 2\% on $\Delta y_t$~\cite{deBlas:2018mhx}.

\begin{acknowledgments}
The authors are grateful to Ambresh Shivaji for useful discussions and for providing some independent checks. A.M. thanks the Department of Physics at Princeton University for hospitality during the completion of this work. The work of S.B. and A.M. is supported by the European Research Council Consolidator Grant NNLOforLHC2. The work of A.M. is also supported by the UK STFC grants ST/L002760/1 and ST/K004883/1.
\end{acknowledgments}

\end{document}